\documentclass{elsart}
\usepackage{graphicx}
\usepackage{amsmath}
\usepackage{amssymb}

\begin{document}

\begin{frontmatter}

\title{Superlattice properties of semiconductor nanohelices
in a transverse electric field}

\author[address1]{O.~V.~Kibis\thanksref{thank1}}
and
\author[address2]{M.~E.~Portnoi}

\address[address1]{Department of Applied and Theoretical Physics,
Novosibirsk State Technical University, Novosibirsk 630092,
Russia}

\address[address2]{School of Physics, University of Exeter, Exeter EX4 4QL,
United Kingdom}

\thanks[thank1]{
Corresponding author.
E-mail: Oleg.Kibis@nstu.ru}

\begin{abstract}
A charge carrier confined in a quasi-one-dimensional semiconductor
helical nanostructure in the presence of an electric field normal to
the axis of the helix is subjected to a periodic potential
proportional to the strength of the field and the helix radius. As a
result, electronic properties of such nanohelices are similar to
those of semiconductor superlattices with parameters controlled by
the applied field. These properties include Bragg scattering of
charge carriers by a periodic potential, which results in energy gap
opening at the edge of the superlattice Brillouin zone. This
provides an opportunity for creating a new class of tunable
high-frequency devices based on semiconductor nanohelices.
\end{abstract}

\begin{keyword}
nanohelices \sep superlattices \sep quantum wires
\PACS 73.21.Cd \sep 73.21.Hb
\end{keyword}
\end{frontmatter}

%[main text]
%\section{Proceedings}

Semiconductor superlattices posses unique physical
properties~\cite{Ivchenko97}, which form a basis for a broad range
of nanoelectronic devices. Existing superlattices are usually based
on multi-layer semiconductor heterostructures. The parameters of
their periodic potential are therefore defined by the growth
conditions and cannot be manipulated subsequently. For many
applications it is highly desirable to develop a new kind of
superlattice in which the periodic potential parameters may be
altered by external fields. This will provide an opportunity for
creating a new class of high-frequency nanodevices with tunable
properties. Here we present a theoretical analysis of a novel type
of electric-field-controlled superlattice based on semiconductor
nanohelices~\cite{Prinz}.

\begin{figure} [htb]
%\vspace{0.5cm}
\begin{center}
\includegraphics[width=12.0cm]{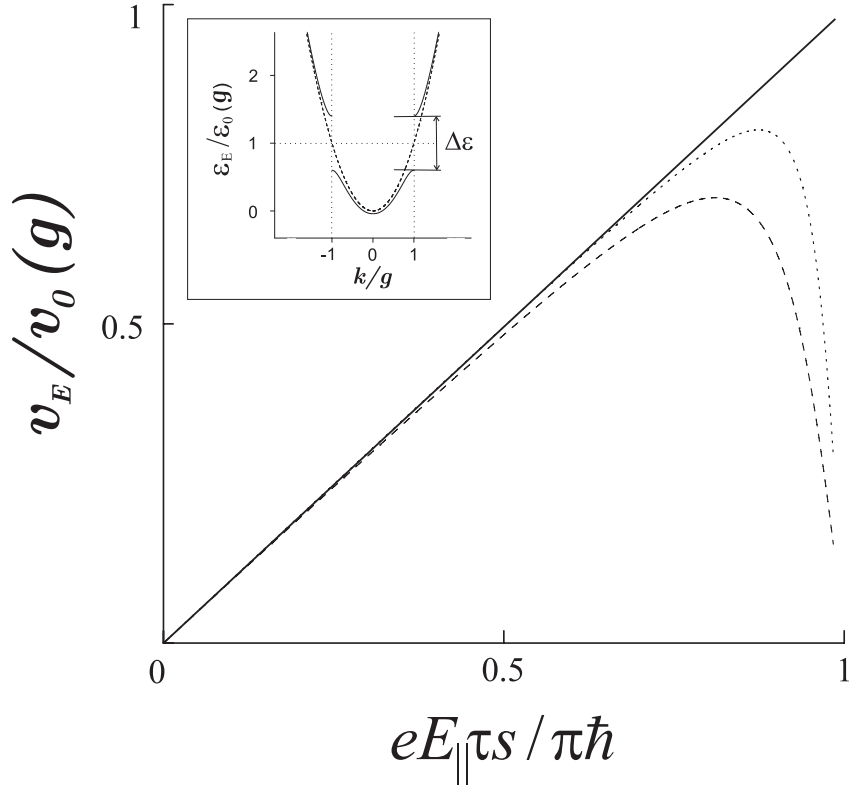}
\end{center}
\caption{ Electron drift velocity in a nanohelix, $v_E$, as a
function of longitudinal electric field, $E_{||}$, directed along
the nanohelix axis: solid line corresponds to $\bar{U}_E=0$, dot
line corresponds to $\bar{U}_E=0.1$, dashed line corresponds to
$\bar{U}_E=0.2$. The inset shows the change of electron energy
spectrum due to transverse electric field: solid line shows the
spectrum in the presence of the field, dashed line --- without
the field.}
\label{fig1}
\end{figure}

Let us consider a quantum wire curved into the shape of a helix with
radius $R$ and pitch $s$ in the presence of an external transverse
electric field $E_{\perp}$, normal to the axis of the helix. Due to
this field, the potential energy of an electron in the helix is
given by
\begin{equation}\label{Uhelix}
U(x)=eE_\perp R\cos\left(2\pi x/l_0\right),
\end{equation}
where $e$ is the electron charge, $x$ is the electron coordinate
along the one-dimensional conductor and $l_0 = \sqrt{4\pi^2R^2+s^2}$
is the length of a single coil of the helix. Evidently, the
potential energy \eqref{Uhelix} is periodic with respect to the
electron coordinate $x$, with period $l_0$, which is significantly
larger than the interatomic distance in the wire. As a result, the
nanohelix acquires typical superlattice properties. In the framework
of the effective mass approximation, the energy spectrum
$\varepsilon_E$ of an electron in a helix in a transverse electric
field is obtained from the Schr\"{o}dinger equation
\begin{equation}\label{Schrhelix}
-\frac{\hbar^2}{2m}\frac{d^2}{dx^2}\psi_E+U(x)\psi_E=
\varepsilon_E\psi_E,
\end{equation}
where $m$ is the effective mass of the electron. The solutions
$\psi_E$ of Eq.~\eqref{Schrhelix} are known to be Mathieu functions
\cite{GradshteynRyzhik07}. However, despite the exact solutions
being known, it is impossible to write the energy spectrum
$\varepsilon_E$ in analytic form as a function of the electron
wavenumber $k$ along the helical line. Since the dispersion
$\varepsilon_E(k)$ determines the main electrophysical parameters of
a superlattice, it is necessary to find it explicitly. We will
therefore use an approximate method of solving
Eq.~\eqref{Schrhelix}. The wave function $\psi_E$, which satisfies
the Schr\"{o}dinger equation with periodic potential~\eqref{Uhelix},
can be written in the Bloch form
\begin{equation}\label{3}
\psi_E=e^{ikx}\sum_{n=-\infty}^{\infty}b_{n}(k)e^{i(2ngx)},
\end{equation}
where $g=\pi/l_0$ is a half of the superlattice Brillouin zone.
Substituting the Bloch function~\eqref{3} in the Schr\"{o}dinger
equation~\eqref{Schrhelix}, we obtain the infinite system of linear
algebraic equations
\begin{equation}\label{Linear}
\left[\varepsilon_0\left(k+2ng\right)
-{\varepsilon_E(k)}\right]b_{n}(k)
+U_E\left[b_{n-1}(k)+b_{n+1}(k)\right]=0,
\end{equation}
where $n$ is the electron subband number,
$\varepsilon_0(k)={\hbar^2k^2}/{2m}$ is the unperturbed energy of an
electron in a nanohelix in the absence of an external electric
field, and $U_E=eE_\perp R/2$ is the characteristic energy of
interaction of the electron with the electric field. The system of
equations~\eqref{Linear} clearly shows that a nanohelix in a
transverse electric field represents a superlattice with the
parameters controlled by the applied field $E_\perp$. Particularly,
the Bragg scattering of the charge carrier by the periodic
potential~\eqref{Uhelix} results in an energy gap opening at the
edge of the superlattice Brillouin zone (see the inset to
Fig.~\ref{fig1}). It should be noted that the similar situation
takes place in carbon nanotubes with helical crystal structure in
the presence of a transverse electric field \cite{Kibis05}.

In what follows we will be mostly interested in the lowest branch of
the energy spectrum. For weak enough electric fields, satisfying the
condition $\bar{U}_E=U_E/\varepsilon_0(g)\ll1$, when solving the
system of equations \eqref{Linear} for this branch, it is sufficient
to take into account the admixture by the electric field of three
electron states lowest in energy, neglecting the contribution of all
other states. In this approximation the system of equations
\eqref{Linear}, which defines the electron energy spectrum, is reduced
to just three equations, from which
the dimensionless electron energy for the ground subband can be
written within the first Brillouin zone as
\begin{equation}\label{Enhelix}
\bar{\varepsilon}_E(\bar{k})=-(2/3)\sqrt{16+48\bar{k}^2
+6\bar{U}_E^2}\cos(\alpha/3-\pi/3)+\bar{k}^2+8/3,
\end{equation}
where $\bar{\varepsilon}_E=\varepsilon_E/\varepsilon_0(g)$,
$\bar{k}=k/g$, and
$$\cos\alpha=-\frac{1-9\bar{k}^2+9\bar{U}_E^2/16}{(1+3\bar{k}^2+
3\bar{U}_E^2/8)^{3/2}}$$ with $0\leq\alpha\leq\pi$. In the same
approximation, the energy gap between the ground subband and the
next one is given by
\begin{equation}\label{Gap}
\Delta\varepsilon=eE_\perp R.
\end{equation}

Superlattices based on semiconductor nanohelices can be used in
electronic devices, which are currently based on conventional
semiconductor superlattices, e.g., quantum cascade lasers
\cite{Faist94}, high-frequency generators and amplifiers
\cite{Esaki70}, etc. The opportunity to control parameters of these
devices by the applied electric field, $E_\perp$, provides
additional flexibility. As an example, let us consider a
nanohelix-based superlattice subjected to an additional longitudinal
electric field, $E_{||}$, along the axis of the helix. We assume
that in the absence of the longitudinal field the electron gas fills
the states near the bottom of the ground subband. In the limit of
weak longitudinal field, the typical change of the electron wave
vector in the field, $\Delta k=eE_{||}\tau s/\hbar l_0$, is less
than the width of superlattice Brillouin zone $2\pi/l_0$ (here
$\tau$ is the electron momentum relaxation time). Then the electric
current in the nanohelix is $j=env_E$, where $n$ is the electron
density, and the drift velocity of electrons, $v_E$, is described by
the Esaki-Tsu formula \cite{Esaki70}
\begin{equation}\label{vE}
v_E=\int_0^\infty{e^{-t/\tau} \frac{d v(eE_{||} t s/\hbar
l_0)}{dt}dt} \approx v(\Delta k).
\end{equation}
Here $v(k)=(1/\hbar)(\partial\varepsilon_E(k)/\partial k)$ is the
electron velocity in the ground subband,
\begin{equation}\label{v}
v(\bar{k})=v_0(g) \frac{3\bar{\varepsilon}_E^2\bar{k}-
6\bar{\varepsilon}_E\bar{k}^3+3\bar{k}^5-16\bar{k}^3+
16\bar{k}-2\bar{U}_E^2\bar{k}}{3\bar{\varepsilon}_E^2
-2\bar{\varepsilon}_E(3\bar{k}^2+8)+3\bar{k}^4+16-2\bar{U}_E^2},
\end{equation}
where $\bar{\varepsilon}_E$ is given by Eq.~\eqref{Enhelix}, and
$v_0(g)=\hbar g/m$ is the electron velocity at the edge of the
superlattice Brillouin zone in the absence of the transverse
electric field $E_\perp$. From the drift velocity plot shown in
Fig.~\ref{fig1} one can see that for $E_\perp\neq 0$ and for high
enough longitudinal electric fields, the electron drift velocity
$v_E$ decreases with increasing electric field $E_{||}$. This means
the appearance of negative differential conductance with the
threshold controlled by transverse electric field. This effect is
caused by electron heating in the longitudinal electric field
resulting in filling electron states with negative effective mass
near the edge of the superlattice Brillouin zone. It should be noted
that the further increase of electric field $E_{||}$ will result in
the Zener tunneling breakdown of the energy gap \eqref{Gap}, and the
current $j$ will again increase with increasing $E_{||}$. As a
consequence, the current-voltage characteristic of a nanohelix will
be of $N$-type, which is similar to tunnel and Gunn diodes. In the
same fashion as these diodes, the nanohelices might be used in
amplifiers and generators.

The effect of negative effective mass near the edge of the
superlattice Brillouin zone also leads to an efficient frequency
multiplication in nanohelices. Let us consider an alternating
electric field along the nanohelice axis, $E_{||}(t)=E_0\cos\omega
t$. Then the electric current in the nanohelix, as a function of
time, can be written as the Fourier series
\begin{equation}\label{Fourier}
j(t)=\sum_{m=-\infty}^{\infty} j_me^{-im\omega t}.
\end{equation}
If $\omega\tau\gg 1$, the coefficients in Eq. \eqref{Fourier} are
given by
\begin{equation}\label{harmonics}
j_m=\frac{en}{2\pi}\int_{-\pi}^{\pi}v\left(\frac{eE_0
s}{\hbar\omega l_0}\sin\omega t\right)e^{im\omega t}d(\omega t).
\end{equation}
From Eq. \eqref{v} it follows that the odd harmonics of the base
frequency $\omega$ should appear in the response current. This
effect can be used in frequency multipliers based on nanohelices.

In the case of weak scattering, when $\Delta k\gg 2\pi/l_0$, the
longitudinal electric field results in Bloch oscillations
 of electrons \cite{Esaki70} with the frequency
$\omega_E=eE_{||}s/\hbar$. For a typical nanohelix pitch of $s\sim
10$nm and electric field $E_{||}\sim 10^4$ V/cm, the oscillation
frequency $\omega_E\sim 10^{12}$ c$^{-1}$ lies in the terahertz
range. This opens the possibility to use nanohelices \cite{Prinz}
for resolving current challenges in terahertz optoelectronics
\cite{THz}.

This work is supported by the INTAS Foundation (Grant
05-1000008-7801), the Royal Society (UK), and the Russian
Foundation for Basic Research (Grant 06-02-16005).

\end{document}